\documentclass[runningheads,a4paper]{llncs}

\usepackage[
    style=ieee,
    doi=false,
    isbn=false,
    url=false,
    eprint=false,
    backend=bibtex,
    natbib=true
    ]{biblatex}

\renewcommand{\cite}{\parencite}

\usepackage{perl_acronyms}
\usepackage{graphicx}
\usepackage{amsmath,amsfonts,bm}
\usepackage{amssymb}
\usepackage{subfigure}
\usepackage{url}
\usepackage{tikz} 
\usetikzlibrary{positioning}
\usetikzlibrary{calc}
\usepackage[inline]{enumitem}
\usepackage{floatrow}

\usepackage{courier}

\newfloatcommand{capbtabbox}{table}[][\FBwidth]

\usepackage{blindtext}

\urldef{\mailsc}\path|{eiscar, satulya, goumas, mattjr}@umich.edu|

\title{
Low cost underwater acoustic localization}

\titlerunning{}

\author{Eduardo Iscar \and Atulya Shree \and Nicholas Goumas \and Matthew Johnson-Roberson}

\institute{University of Michigan\\Department of Naval Architecture and Marine Engineering\\
2600 Draper Drive, Ann Arbor, MI 48109\\
\mailsc
}

\toctitle{Low cost underwater acoustic localization}
\authorrunning{E. Iscar, A. Shree, N. Goumas and M. Johnson-Roberson }

\begin{document}

\maketitle

\begin{abstract}
Over the course of the last decade, the cost of marine robotic platforms has significantly decreased. In part this has lowered the barriers to entry of exploring and monitoring larger areas of the earth's oceans. However, these advances have been mostly focused on \acp{ASV} or shallow water \acp{AUV}. One of the main drivers for high cost in the deep water domain is the challenge of localizing such vehicles using acoustics. A low cost \ac{OWTT} underwater ranging system is proposed to assist in localizing deep water submersibles. The system consists of location aware anchor buoys at the surface and underwater nodes. This paper presents a comparison of methods together with details on the physical implementation to allow its integration into a deep sea micro AUV currently in development. Additional simulation results show error
reductions by a factor of three.
\end{abstract}

\section{Introduction}
\begin{figure}[]
\centering
\includegraphics[width=0.55\textwidth ]{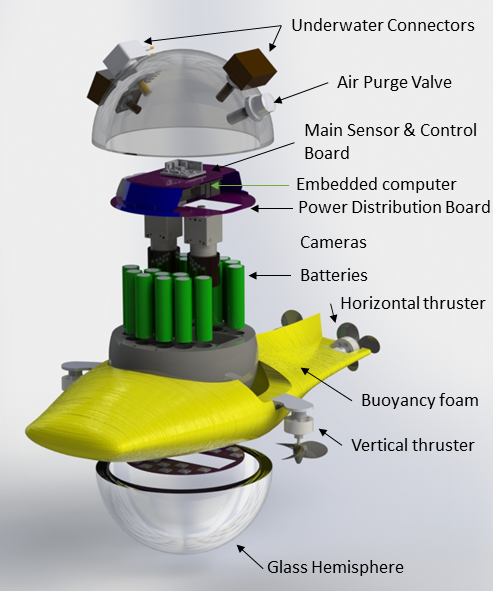}
\caption[]{Exploded view of the DS$\mu$AUV}
\label{fig:auv}
\vspace{-2mm}
\end{figure}

The localization of underwater targets has been one of the goals of underwater acoustics since its inception in the early 20th century. More recently, the rise of underwater sensor networks used in environmental monitoring, natural disaster prevention or oil drilling operations has further boosted the need for accurate underwater localization. In the case of \acp{AUV}, the lack of \ac{GPS} makes the use of acoustic sensing  necessary. Sensors  such as the  \ac{DVL} or \ac{LBL} localization allow the vehicle to compute its position underwater. Camera based optical methods such as visual odometry have also been successfully applied underwater~\cite{wirth2013visual}, but, similarly to other dead reckoning methods, suffer from error accumulation over time. We have developed a novel, deep sea capable, low cost micro \ac{AUV} equipped with cameras, an \ac{IMU} and a depth sensor. This platform is targeted for deep sea optical mapping. Figure~\ref{fig:auv} shows an exploded view of the $\mu$AUV with its main components labeled. This platform was designed to help gather high resolution optical maps of the deep ocean. While high accuracy maps of terrestrial terrain are widely available at centimeter resolution, ocean bathymetry is much coarser with maps at $1m$ resolution per pixel only available for 0.05\% of the ocean~\cite{copley2007}. Medium resolution sonar based maps are available for scattered areas, and do not exceed grid resolutions of a hundred meters. The developed $\mu$\ac{AUV} is able to  provide photographic mosaics and accurate depth measurements of areas of interest. In order to georeference these maps  accurate vehicle position estimates are needed.  Although \ac{COTS} localization solutions are available, their cost and size makes them unsuited for use in the proposed \ac{AUV}. \\
This paper has two major contributions:  First we present the initial development of an acoustic localization system based exclusively on \ac{OWTT} range measurements for low cost \acp{AUV}. Then we show simulation results highlighting the localization performance improvements due to the ranging information. The  paper is structured as follows: In Section~\ref{sec:background} related work in the field of acoustic localization and sensor data fusion is presented. Section~\ref{sec:technical} introduces the developed hardware and software to perform acoustic range estimation. Section~\ref{sec:datafusion} presents the sensor fusion framework and shows simulation results. Finally, Section \ref{sec:conclusion} presents our conclusions and future work.
\section{Background}
\label{sec:background}
Underwater acoustic localization methods can be divided into two main groups: Range based and range free methods. Range free methods provide very coarse position estimates based on hop counting~\cite{wong2005multihop} or Point in Triangle methods~\cite{he2003range}, and are thus not suited for mapping purposes. Range based methods rely upon the calculation of the \ac{ToA}, \ac{TDoA}, \ac{TWTT} or \ac{RSSI} to determine the distance between two nodes. \ac{RSSI} methods are specially susceptible  to  multipath and scattering effects~\cite{chandrasekhar2006localization}. As a consequence, \ac{ToA} and \ac{TDoA} are the preferred methods. Successful computation of the \ac{ToA} additionally requires time synchronization between nodes. Cheng et al. \cite{cheng2008silent} proposed the use of \ac{TDoA} from multiple anchor nodes to eliminate the requirement for time synchronization. However, the requirement of fixing nodes to the seafloor make it unsuitable for quick deployment missions. Eustice et al. \cite{eustice2006recent} proposed the use of a nonlinear least squares weighting scheme to estimate the position of a buoy and \ac{AUV} through clock-synchronized \ac{OWTT}.

In the last decade, acoustic localization has been frequently used inside data fusion algorithms like \acp{EKF}~\cite{webster2009preliminary} or \acp{PF}~\cite{ko2012particle}. \acp{PF} show superior performance in the presence of highly non-linear measurement models or high variance \cite{thrun2005probabilistic}, but require a high number of particles (and processing power) as the dimensionality of the state space increases. 
While most methods presented so far involve a single vehicle with statically deployed anchor nodes or buoys, cooperative approaches with multiple vehicles have also been proposed to solve the localization problem. Such scenarios usually involve at least one vehicle equipped with either high-accuracy navigation sensors or \ac{GPS} when operating at the sea surface. The position of all vehicles can then be estimated through Centralized Extended Kalman filters~\cite{webster2012advances} or Decentralized Extended Information Filters~\cite{webster2013decentralized}.
The requirement of not only time-synchronized clocks but also of transmitting data through the acoustic channel imposes higher system complexity and makes these methods incompatible with low-cost underwater swarms. 

A more detailed survey of underwater localization methods can be found in the works of Chandrasekhar et al.~\cite{chandrasekhar2006localization} and Tan et al.~\cite{tan2011survey}.

\section{Acoustic Range Estimation}
\label{sec:technical}

\begin{figure}
\centering
\includegraphics[width=0.99\textwidth ]{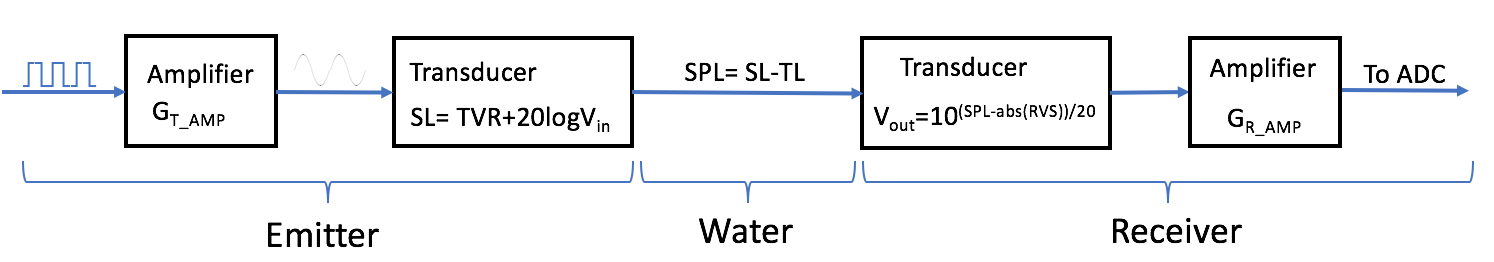}
\caption[]{Acoustic ranging system flowchart: The electronic amplifiers are discussed in detail in Section~\ref{sec:amplifiers}, while the transducer design and characterization is shown in Section~\ref{sec:transducers}}
\label{fig:system_flowchart}
\vspace{-4mm}
\end{figure}

Although acoustic positioning is  frequently combined with underwater data transmission, we decided to limit the system functionality to range estimation in order to reduce system complexity and increase reliability. The adopted solution was \ac{OWTT}, which enables passive listening on the vehicles, reducing the energy consumed and thus increasing vehicle endurance.  However, the price for the low energy footprint is the requirement for low drift synchronized clocks between the transmitter and all receivers. 
Figure~\ref{fig:system_flowchart} illustrates the process of range estimation through \ac{OWTT}. A detailed discussion of the transducers is presented in Section~\ref{sec:transducers}, while both transmitter and receiver amplifier designs will be discussed in Section~\ref{sec:amplifiers}. On the left of Figure~\ref{fig:system_flowchart}, the transmitter generates a tonal pulse at a specified time interval. The pulse is bandpass filtered and amplified to drive the transducer, which converts the electrical energy into acoustical pressure waves according to Equation~\ref{eq:transducer_transmit}, where the signal voltage $V_{RMS}$ acts on a transducer of a given \ac{TVR} to generate the output Sound Level(SL) at $1m$ distance given by:
\begin{equation}
        SL (dB) = TVR + 20\log{V_{RMS}}
    \label{eq:transducer_transmit}
\end{equation}

The acoustic wave attenuates as it travels through the water. The two main components of \ac{TL} are spreading and attenuation. For deep water, spreading is assumed to be spherical, which leads to a spreading loss of $20\log{R}$, where $R$ is the distance to the transmitter. Attenuation is a frequency dependent phenomenon, frequently modeled as $ \alpha (f)  R$, where $\alpha(f)$ is the absorption coefficient and $R$ the range from the transmitter. This leads to the following expression for \ac{TL}:

\begin{equation}
        TL (dB) = 20\log{R} + \alpha(f)R
    \label{eq:transmission_loss}
\end{equation}

At the receiving end, the acoustic signal is converted back into electrical energy by the receiving transducer. Equation~\ref{eq:transducer_receive} relates the \ac{SPL} at $1m$ of the receiver with the electrical signal amplitude in $V_{RMS}$ and the transducer \ac{RVS}: 

\begin{equation}
        SPL = \lvert \text{RVS}\rvert + 20\log{V_{RMS}}
        \label{eq:transducer_receive}
\end{equation}

The electrical signals generated by the hydrophones have very small amplitudes and thus require amplification before interfacing with the \ac{ADC}. After being digitized, the use of correlation and a \ac{SDFT} allows computation of the time delay. 

\subsection{Acoustic Transducers}
\label{sec:transducers}
One of the main cost drivers for underwater acoustic systems are transducers. In the following section, we present a set of four different transducer designs and their relative performance. The main element of the transducers is the piezoelectric cylinder. Piezoelectric crystals show a linear relation between mechanical strain and electrical field~\cite{sherman2007transducers}. This property is used to create mechanical vibrations that generate acoustical pressure waves. 
We build upon the work of Benson et al.~\cite{benson2010design} in the selection of piezoelectric rings with a resonance frequency of 43kHz~\cite{ceramictransducer}. Higher frequencies attenuate faster in water, making lower frequencies better suited for long range. Additionally, lower frequencies impose less stringent requirements on the data acquisition and processing hardware. 

\begin{figure}
\centering
\subfigure[\label{fig:transducer_assembly}  Exploded transducer view ]{
\includegraphics[width=0.30\textwidth ]{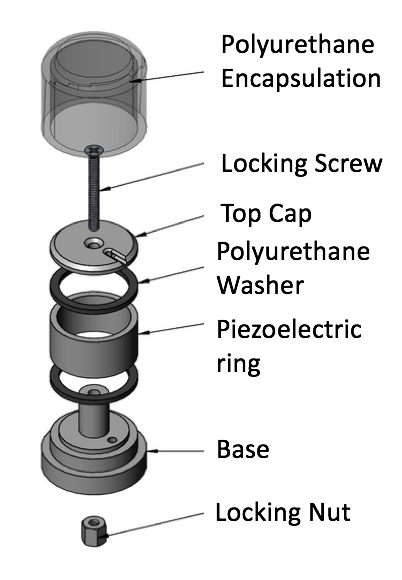}
}
\subfigure[\label{fig:transducer_dl4}  Single ring transducer ]{
\includegraphics[width=0.27\textwidth ]{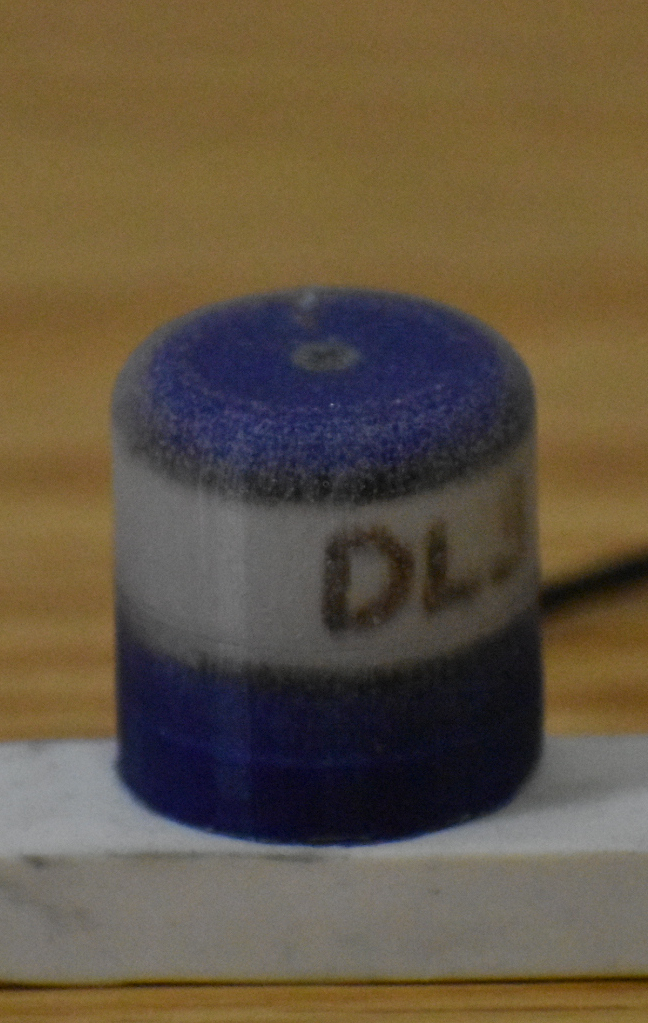}
}
\subfigure[\label{fig:transducer_dl7} Double stack transducer]{
\includegraphics[width=0.24\textwidth]{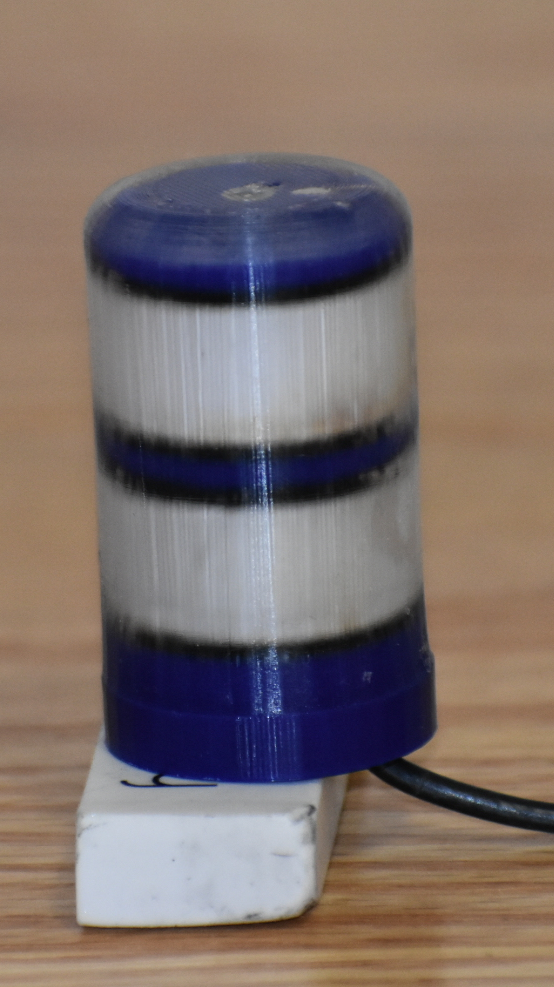}
}
\caption[]{Custom made transducers}
\label{fig:customtrans}
\vspace{-4mm}
\end{figure}

In Figure~\ref{fig:transducer_assembly} an exploded view of the transducer design can be observed. Four main pieces form the transducer body:
\begin{enumerate*}[label=(\roman*)]
  \item a cylindrical base with mounting holes to allow the attachment of the transducer to a flat surface, as well as pass-through holes for the transducer cable and locking screw, 
  \item a set of two polyurethane washers to distribute compression forces evenly over the ring top and bottom faces.
  \item the piezoelectric ring, soldered to a shielded coaxial cable and
  \item a top end cap disc.
\end{enumerate*}
 A screw locks the elements together. Both the transducer base and end cap have been prototyped using 3D printing, but could easily be machined out of ABS or other materials if needed. Waterproofing and electrical isolation is achieved by potting the transducer assembly in polyurethane resin. We chose an optically clear resin to allow easy inspection of the finished assembly. The resin density  (1060$kg/m^3$) is very similar to salt water density and helps transfer the acoustic waves from the polyurethane to the water. Figure \ref{fig:transducer_dl4} shows the finished transducer. Detailed instructions and material lists are provided under \url{https://bitbucket.org/eiscar/uar/wiki/Home}.

Four different configurations were fabricated to investigate the effect of different cylinder core materials, as well as the potential gains by chaining multiple piezoelectric rings. The tested configurations were:
\begin{enumerate*}[label=(\arabic*)]
  \item a fully potted core,
  \item a fully potted core with an additional layer of cork around the interior face of the cylinder,
  \item an air gaped ring and
  \item a stack of two air gaped rings.
\end{enumerate*}
In the fully potted transducer, the core of the piezoelectric oscillator ring is filled with the same  polyurethane used for the exterior encapsulation. According to~\cite{sherman2007transducers}, this increases the radial stiffness and reduces effective coupling. This effect can be reduced by  the addition of a layer of cork to the interior face that isolates the polyurethane from the electrode. Air cavities were created for both (iii) and (iv). The connection of the ceramic rings in series is expected to generate higher sensitivities. Figure~\ref{fig:transducer_dl7} shows the double ring transducer. 

\begin{figure}
\centering
\subfigure[\label{fig:transducer_impedance_pot} Receive Sensitivity]{
\includegraphics[width=0.6\textwidth]{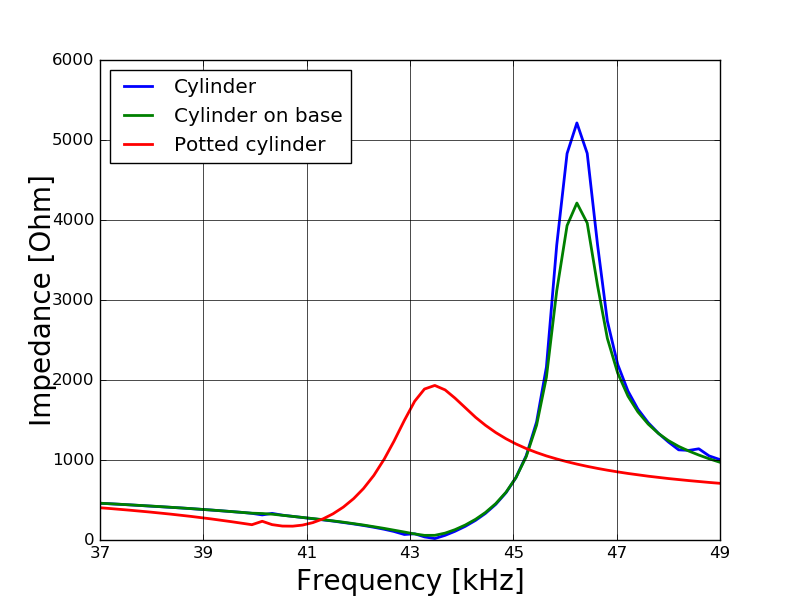}
}
\subfigure[\label{fig:transducer_impedance_com} Comparison of transducer impedance curve]{
\includegraphics[width=0.6\textwidth ]{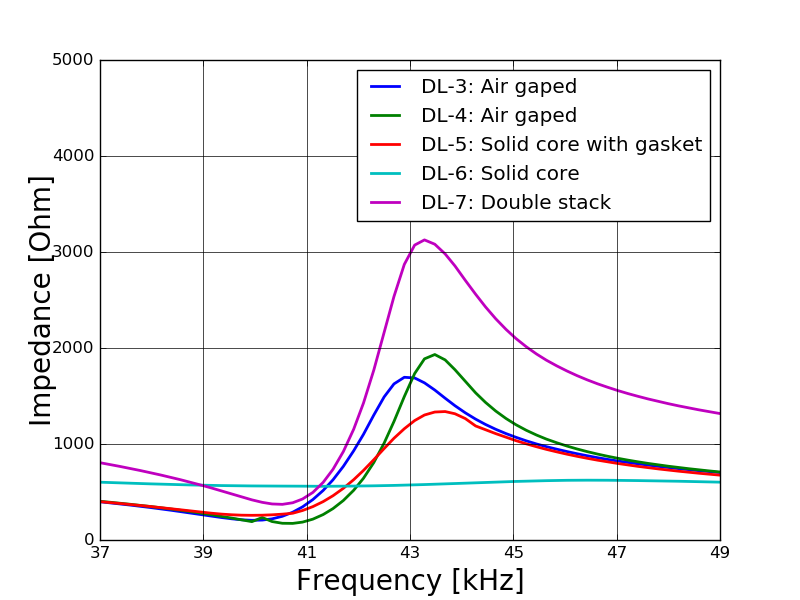}
}
\caption[]{\ref{fig:transducer_impedance_pot} shows the impedance response of the DL-4 transducer during the different transducer assembly steps. Resonance frequency is reduced as oscillator motions are constrained by the support structure and potting polyurethane and needs to be taken into account when selecting the oscillator for a target frequency. \ref{fig:transducer_impedance_com} compares the transducer impedance of the different configurations, showing minor effects on the final resonance frequency but large  variation of the impedance curve amplitude.}
\label{fig:trans_impedance}
\end{figure}

Figure \ref{fig:transducer_impedance_pot} shows the plot of the transducer impedance against frequency at the different stages of assembly. Impedance oscillation amplitude decreases as the oscillator is constrained. The resonant frequency, which occurs at the minimum of the impedance plot, shifts from 43kHz to around 40kHz. Figure~\ref{fig:transducer_impedance_com} compares the impedance curves of the finished transducers for the different configurations. While each different transducer configuration had little impact on resonant frequency, the amplitude of the impedance oscillation greatly varies between each of the tested prototypes.

In order to characterize the hydrophones receiver response we obtained the \ac{RVS} frequency curve by calibrating with respect to an  Aquarian Scientific AS-1 reference hydrophone. The main characteristics of the AS-1 are shown in Table~\ref{tab:as-1}. At \$400, they are relatively low cost, but provide the acoustic calibration to serve as a baseline. By measuring the response of both transducers to the same acoustic signal, the free field sensitivity of the unknown transducer can be determined using Equation~\ref{eq:cal}:  

\begin{equation}
    \lvert \text{RVS}_{cal}(f)\rvert = \lvert \text{RVS}_{ref}(f)\rvert- (G_{ref}(f)-G_{cal}(f)) + 20\cdot\log{\frac{V_{ref}(f)}{V_{cal}(f)}}
    \label{eq:cal}
\end{equation}
where $G_{ref}$ is the gain of the amplifier connected to the reference hydrophone, $G_{cal}$ is the gain of the amplifier connected to the transducer being characterized and $V_{ref}$ and $V_{cal}$ are the RMS voltage levels at the output of the receiver.

\begin{figure}
\centering   
\includegraphics[width=0.6\textwidth ]{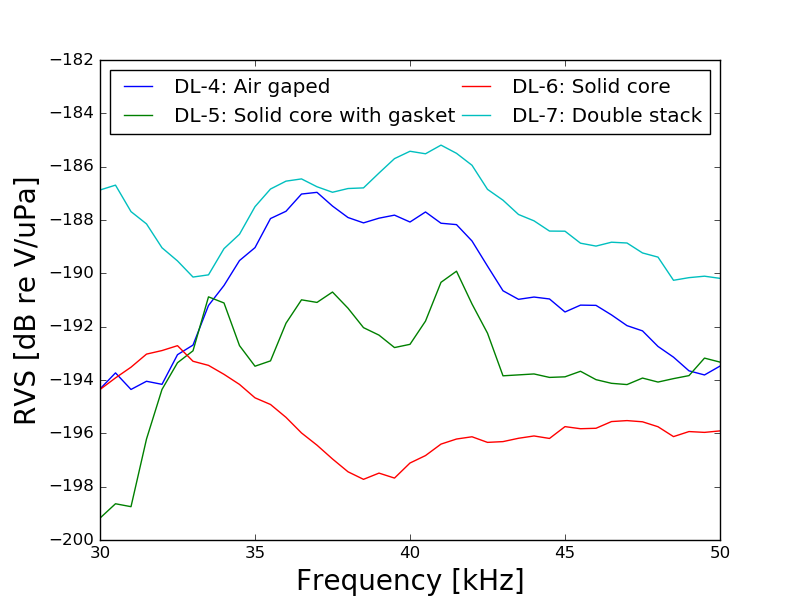}
{\caption{\label{fig:transducer_RVS} Transducer Receiving Voltage Sensitivity. The graph shows how the highest performance is achieved by using the double oscillator ring stack.}
    }

\end{figure}

Figure \ref{fig:transducer_RVS}  shows the receive sensitivity curves for the four transducer designs in the frequencies of interest. As expected, the double ring transducer DL-7 showed the highest sensitivities. The air-core DL-4 transducer was the most sensitive of the single ring configurations. Also noteworthy is the improvement achieved by adding the additional cork gasket when compared to  the solid core. The performance correlates with the amplitude of the impedance curves shown in Figure~\ref{fig:transducer_impedance_com}.

\begin{table}
\begin{tabular}{lc}
    \textbf{Parameter} & \textbf{Value} \\ \hline
    Linear Range & 1Hz to 100kHz  \\
    Receiving Sensitivity & -208dB re~$1V/\mu Pa$ \\
    Transmitting Sensitivity & 140 dB re~$1\mu Pa/V$ \\
    Maximum Input Voltage & 150Vpp \\ 
    Nominal Capacitance & 5.4nF \\
    Operating depth & 200m 
    \end{tabular}
\vspace{7mm}
\caption{\label{tab:as-1}AS-1 Hydrophone characteristics.}
\end{table}

\begin{figure}
\centering
\subfigure[\label{fig:trans_bode} Transmitter Bode Plot]{
\includegraphics[width=0.6\textwidth ]{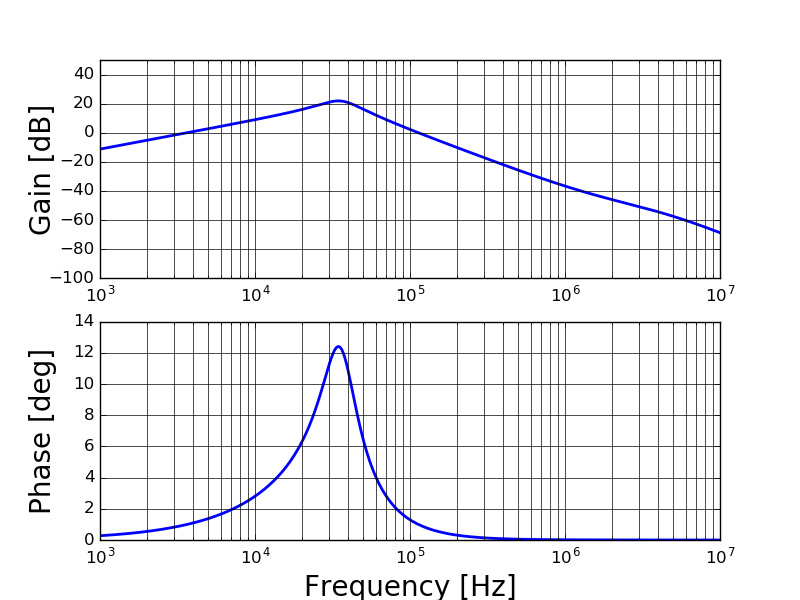}
}
\subfigure[\label{fig:rec_bode} Receiver Bode Plot]{
\includegraphics[width=0.6\textwidth]{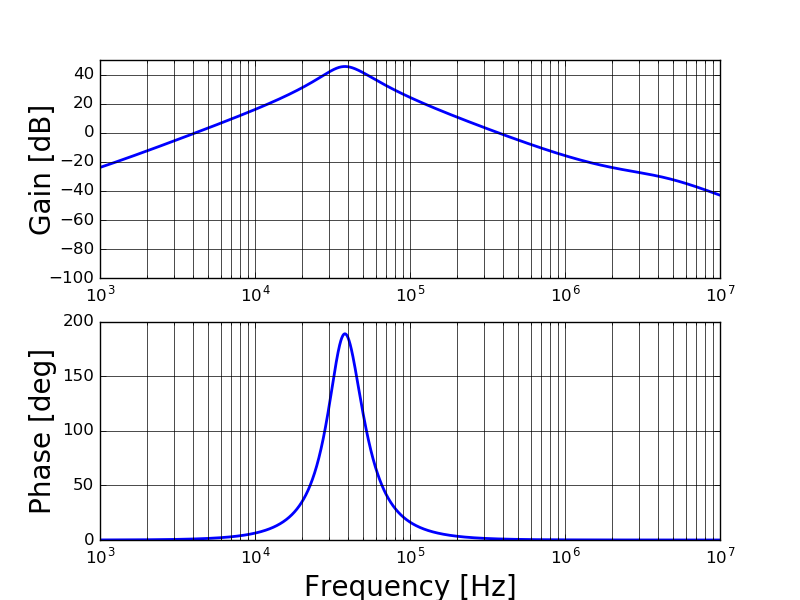}
}
\caption[]{Bode plots for both the transmitter and receiver amplifier boards, showing maximum gains at transducer resonant frequency. }
\vspace{-4 mm}
\end{figure}

\subsection{Transmitter and Receiver amplifiers}
\label{sec:amplifiers}
We build upon the work of Garcia~\cite{garcia2010ultrasonic} and Trezzo~\cite{openrov-forum} for the first revision of the transmitter and receiver amplifier designs. The receiver consists of a preamplifier cascaded with a multiple feedback bandpass filter. The gain of both stages combined is 45dB at 40kHz. Figure~\ref{fig:rec_bode} shows the complete Bode plot for the receiver side. A DC offset voltage of 1.6V is applied before sampling to ensure effective use of the analog input range. Over- and under-voltage protection was also added to the output of the amplifier to ensure it did not exceed the limits of the ADC input. 

The transmitting amplifier converts the 0-3.3V logic levels to a $30V_{pp}$ signal capable of driving the transducer. Its bode plot is shown in Figure~\ref{fig:trans_bode}. Additionally, the amplifier also bandpass-filters the signal, effectively enabling us to drive the transducer from a square wave at the microcontroller output.

\subsection{Signal Detection and Delay Estimation}
The received, amplified signal is sampled by the microcontroller at a frequency of 250kHz. Following \cite{shatara2010efficient} a \ac{SDFT} is applied to the digitally sampled data in real time. The \ac{SDFT} presents the advantage of being computationally efficient when interested  in a specific bin and not the full frequency spectrum of the \ac{DFT}. The k-th spectral bin of a \ac{DFT} can be computed as: 
\begin{equation}
    X_k[n]=\sum_{m=0}^{N-1}x[n-m]e^{\frac{-j2\pi km}{N}}, k = \frac{f_0N}{F_s}
\end{equation}


where $f_0$ is the frequency of interest and $F_s$ is the sampling frequency and  $N$  the number of samples in the window.  The \ac{SDFT} can then be iteratively updated as: 
\begin{equation}
        X_k[n] = (x[n] - x[n-N] + X_k[n-1]) e^{\frac{-j2\pi k}{N}}
\end{equation}

The resulting values are squared to eliminate the complex part and compared to a user-selectable threshold. A circular buffer is filled with the samples until the threshold is exceeded. After the trigger event, a second buffer with the sampled data is filled. After the two buffers are full the delay is calculated using a matched filter, estimating the correlation between the transmitted signal and the sampled received data. The maximum of the correlation function indicates the best alignment between the two signals, and together with the sample rate, the time delay can be computed. 

\begin{figure}[]
\centering
\subfigure[\label{fig:sdft} \ac{SDFT} Trigger]{
\includegraphics[width=0.33\textwidth ]{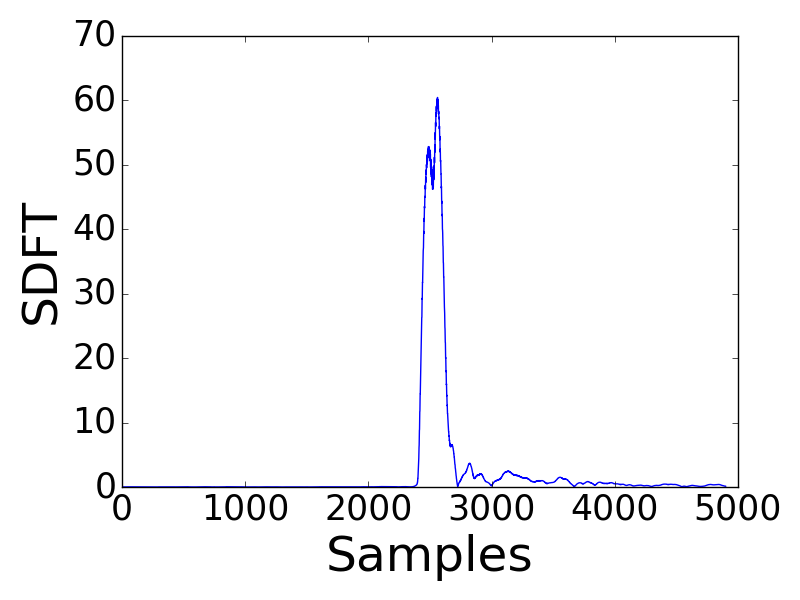}
}
\subfigure[\label{fig:rec_signal} Sampled signal]{
\includegraphics[width=0.29\textwidth]{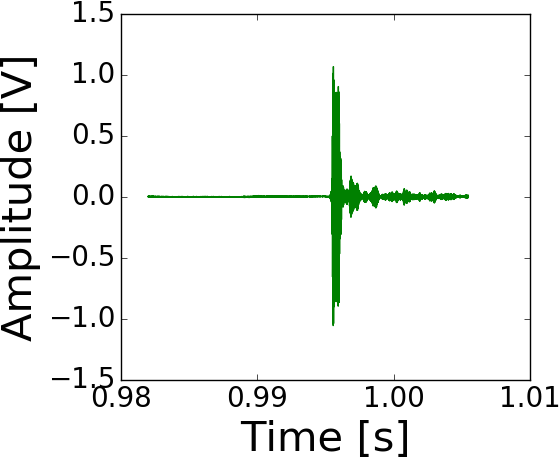}
}
\subfigure[\label{fig:corr} Correlation]{
\includegraphics[width=0.31\textwidth]{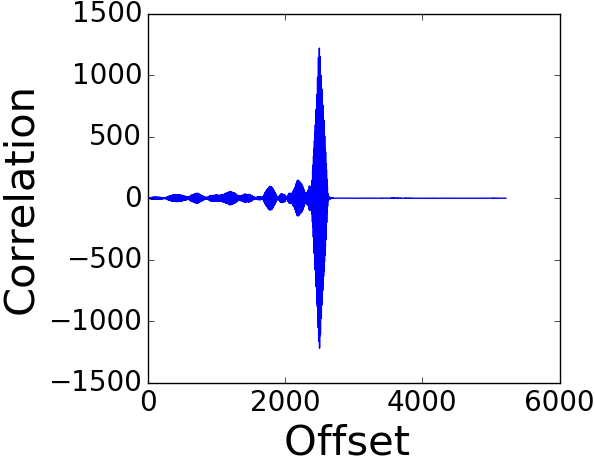}
}
\caption[]{Signal triggering and detection method, illustrating the trigger data in Figure~\ref{fig:sdft}, the acquired data after being digitized at the ADC in Figure~\ref{fig:rec_signal} and the correlation peak to determine receive time offset in Figure~\ref{fig:corr}}
\label{fig:sig_trig_corr}
\end{figure}

Figure~\ref{fig:sig_trig_corr} shows the process of signal detection and delay estimation. In Figure~\ref{fig:sdft} the output of the \ac{SDFT} windowed samples is shown. The captured signal is shown, plotted against time, in Figure~\ref{fig:rec_signal}. Finally, Figure~\ref{fig:corr} shows the output of the correlation function, whose peak is used to estimate the time delay of the aligned signal. 

Once the delay $\Delta t$ of the received signal is obtained, the distance between transducers can be determined as: 
\begin{equation}
        d = \Delta t \cdot c
\end{equation}

where $c$ is the speed of sound in water. 


\subsection{Time of Flight Hardware Experiments  }
\begin{figure}[H]
\centering
\includegraphics[width=0.8\textwidth ]{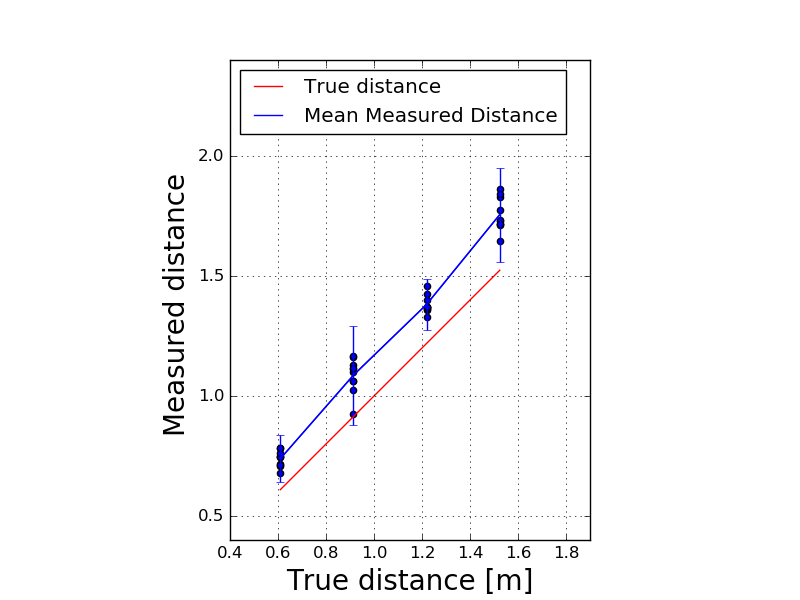}
\caption[]{Range measurement experiments}
\label{fig:experiments}
\end{figure}

We took the development transmitter and receivers to a freshwater pool and tested the \ac{OWTT} method presented in Section \ref{sec:technical} at four different distances. Receiver and transmitter transducers were placed at the same depth in shallow water. For each distance, 10 different measurements were taken. Time synchronization was achieved by using \ac{GPS} on both the transmitter and receiver. In Figure~\ref{fig:experiments} the measurements, in blue dots, are plotted against the true distance shown in red. The error bars show the 3-$\sigma$ deviation error bounds, while the blue line connects the mean of each measurements set.
The experiment results show that the system was able to determine the distance with a high level of precision. The average standard deviation for the considered distances was $3.6cm$. A constant delay is being introduced at an unknown step of the processing pipeline, generating the approximately constant offset of $15cm$ between the blue and red curves. Further investigation into the source of the delay will allow to better characterize and model the system. Compensation can however be simply performed by subtracting the known offset from future measurements.

\section{Simulation of Beacon Based Localization }
\label{sec:datafusion}

In the previous section we have introduced the hardware components required to develop a range measurement sensor using acoustic transducers. In this  section we describe how we can use the acoustic range data along with the vehicles' odometry sensors in order to localize the robot. We run simulations to emulate the sensor data and analyze the accuracy of our localization framework.

\subsection{System Description}
%

\begin{figure}[h]

\centering
\subfigure[\label{fig:system_desc} System Description]{
\includegraphics[width=0.5\textwidth ]{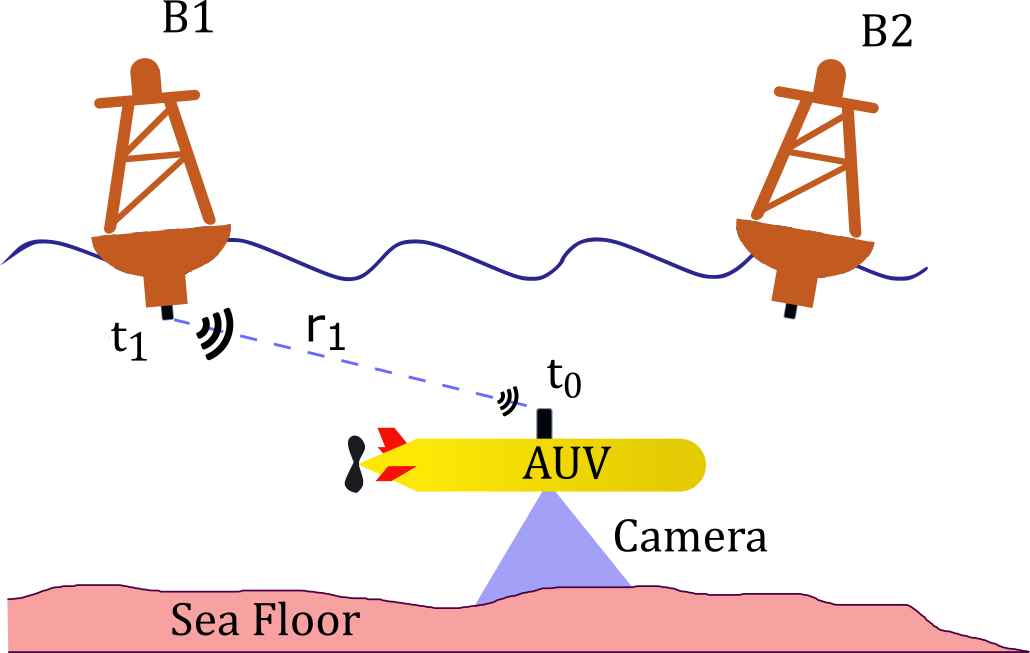}
}
\subfigure[\label{fig:posegraph} Pose graph]{
\includegraphics[width=0.4\textwidth ]{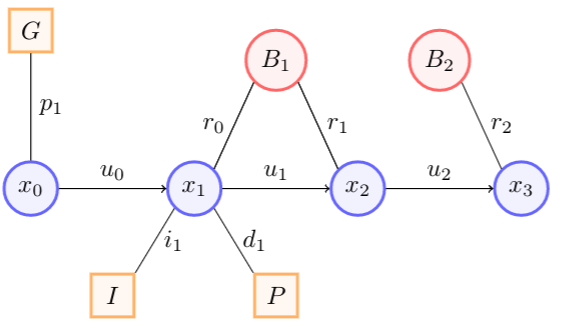}
}
\caption[]{Fig \ref{fig:system_desc} shows the static beacons at the surface and the AUV. Range is measured by \ac{OWTT} of the acoustic ping signal. Fig \ref{fig:posegraph} shows the sample posegraph framework for a small series of measurements}
\end{figure}

The system comprises of 2 stationary buoys at the sea surface. This creates an instrumented workspace bound by the range of the acoustic transducers.  The \ac{AUV} is equipped with a set of cameras which it can use for visual odometry while at the sea floor. At all other times it relies on the dead-reckoning for position estimation. Our system relies on fusion of odometry estimate with the range measurements for accurate localization. Recent approaches have shown pose graphs to be very effective towards solving this problem 
and in the next section we describe how the range measurements are added to a factor graph together with the rest of the vehicle sensors to provide improved localization estimates.
\subsubsection{Pose graph framework}

The problem of localization has been very well studied in the robotics literature and several techniques exist to estimate robot state based on external measurements. Traditional  filtering approaches such as \ac{EKF} and \ac{PF} marginalize past poses, thus discarding information. A pose graph on the other hand incorporates all the measurements and past poses and formulates the localization problem as that of global optimization over all states. While this is computationally more expensive it provides  smoother trajectories and lower errors in presence of non-linearities.

Figure \ref{fig:posegraph} describes a small section of the pose graph that was used in our approach. The robot states($x_i$) and buoy locations $B_i$ are represented as nodes in our graph. Edges between the nodes encode sensor measurements such as a range measurement $(r_i)$ between a state and a buoy. Edges between consecutive states are obtained using the visual odometry sensor $(u_i)$. Prior information about the buoys and information from the GPS is represented as a unary position constraint $(p_1)$ on the nodes. These can also be thought of as global measurements linking the graph to the world coordinate system. The pressure sensor gives a depth measurement $(d_1)$ and the IMU an orientation measurement$(i_1)$ which are again unary constraints on the nodes.

The pose graph framework provides a Maximum a posteriori estimate of the vehicle state that best fits the sensor measurements obtained so far. Since we wish to utilize the framework for real-time position estimation we use the algorithm iSAM2 \cite{kaess2012isam2}. This allows us to incrementally solve the optimization problem in an online manner by variable reordering and fast incremental matrix factorization. 

For more details on solving the above non-linear function as well as additional information on Pose graph techniques, we refer our readers to the work done by Dellaert et al.~\cite{dellaertgtsam}.

\subsubsection{Sensor Characteristics}

The vehicle state is composed of its position expressed in a global \ac{NED} coordinate frame as well as the roll-pitch-yaw orientation angles. The different sensors present on our robot are the acoustic range finders, pressure sensor, \ac{IMU}, \ac{GPS} and a pair of cameras for stereo vision. Sensor noise is based on manufacturer data when available.

The camera images are used for visual odometry by detecting and tracking feature points in the different views and using them for egomotion estimation. 
We heuristically choose the odometry noise characteristics in such a way that that the resulting position error is two orders of magnitude more than a conventional DVL sensor.
The various noise characteristics are described in Table~\ref{tab:sensors}.

\begin{table}[h]
	\centering
	\caption{\label{tab:sensors} Simulated Sensors \& Noise characteristics}
\begin{tabular}{lcc}
	\textbf{Sensor} & \textbf{Frequency (Hz)} & \textbf{Noise StdDev} \\ \hline
	\ac{GPS} & 1 & $0.6m$\\
	AHRS roll, pitch, yaw & 2 & $(2.87^{\circ},2.87^{\circ}, 5.7^{\circ)}$ \\
	Pressure  & 2 & $0.1m$  \\ 
	Acoustic Range & 0.25 & $1m$  \\ 
	Thruster setpoint based Dynamic Model & 10 & $(0.2m, 0.2m, 0.2m)$ \& \\
	 & & $(0.45^{\circ}, 0.45^{\circ}, 0.45^{\circ})$  \\
	Visual Odometry  & 10 & $(0.04m, 0.04m, 0.04m)$ \& \\
	 & & $(1.2^{\circ}, 1.2^{\circ}, 1.2^{\circ})$  \\
\end{tabular}
\vspace{-4mm}

\end{table}


\subsection{Localization Simulation}

%


Simulations are run using  \ac{ROS} in order to evaluate the improvement in localization accuracy introduced due to the addition of acoustic range information. A dynamic model of the robot was computed and used to implement a cascaded PID controller, allowing the vehicle to follow predefined waypoints. We simulated two scenarios that are representative of typical robot operation: 
the first is a dive sequence where the robot launches from the surface and moves to the sea floor, while  the second resembles a survey mission where it follows a lawn mower trajectory over a sea floor. 

\subsubsection{Dive Sequence}In this scenario the robot starts at the water surface and performs a dive along a straight trajectory to a depth of $50m$. The lack of visual features during the descent  makes visual odometry impossible, forcing the use of a dead reckoning system which has significantly higher noise characteristics. 

\begin{figure}[h]

\centering
\subfigure[\label{fig:dive_trajectory} 2D Trajectory plot]{
\includegraphics[width=0.46\textwidth ]{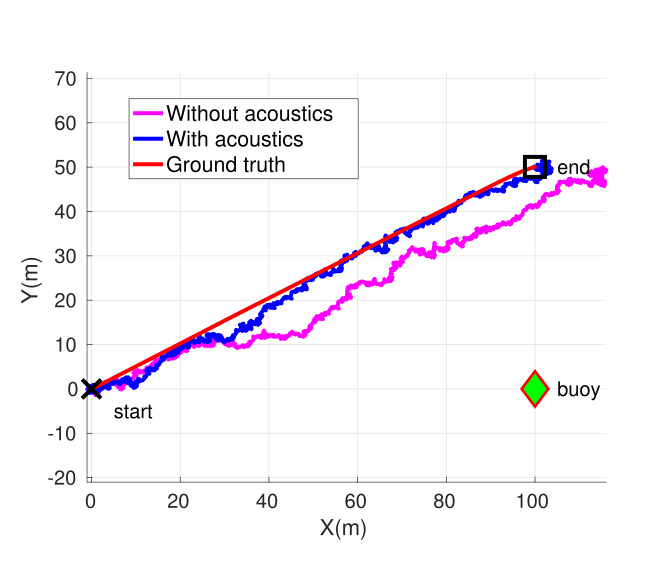}
}
\subfigure[\label{fig:dive_err} Error in X and Y with $ \pm 3\sigma$ bounds]{
\includegraphics[width=0.46\textwidth]{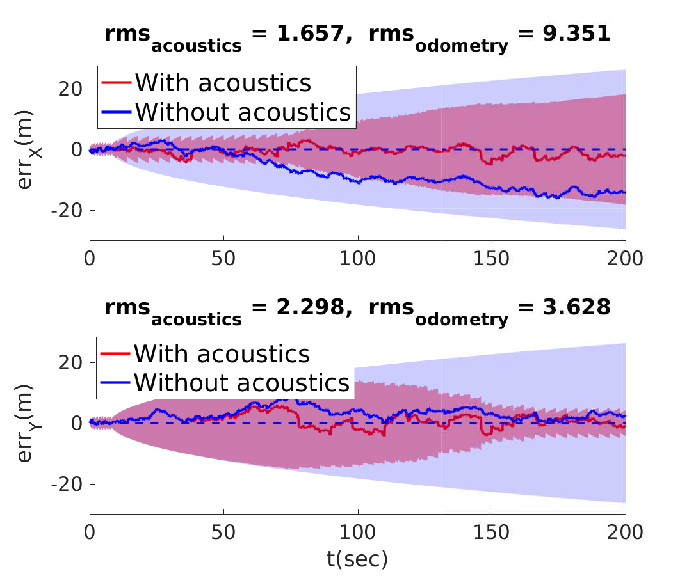}
}
\vspace{-1mm}
\caption[]{Figure~\ref{fig:dive_trajectory}: Improvement in vehicle trajectory with a range sensor. Figure~\ref{fig:dive_err} shows the $3 \sigma$ bounds for error uncertainty over time.}
\vspace{-4mm}
\end{figure} 

Figure \ref{fig:dive_trajectory} shows the trajectory obtained. 
While at the surface, \ac{GPS} measurements bound the uncertainty in the position estimates. During the dive the uncertainty in position due to odometry-only measurements increases monotonically. Using the range sensor demonstrates significant correction capabilities radially from the transmitter, but it cannot correct significant errors along the tangential direction. This can be seen in the error plots of X in Figure~\ref{fig:dive_err}. Along the X axis the uncertainty remains bounded until 70 secs. After that the X direction becomes tangential to the buoy and so the range sensor cannot correct errors along it. The complementary effect can be seen in uncertainty of Y which grows uncontrained initially, but gets tighter bounds for the later half of the path.



\subsubsection{Lawn Sequence}

\begin{figure}[h]
\vspace{-4mm}
\centering
\subfigure[\label{fig: lawn_seq} Buoy placed at $(0, 50, 0)$]{
\includegraphics[width=0.46\textwidth ]{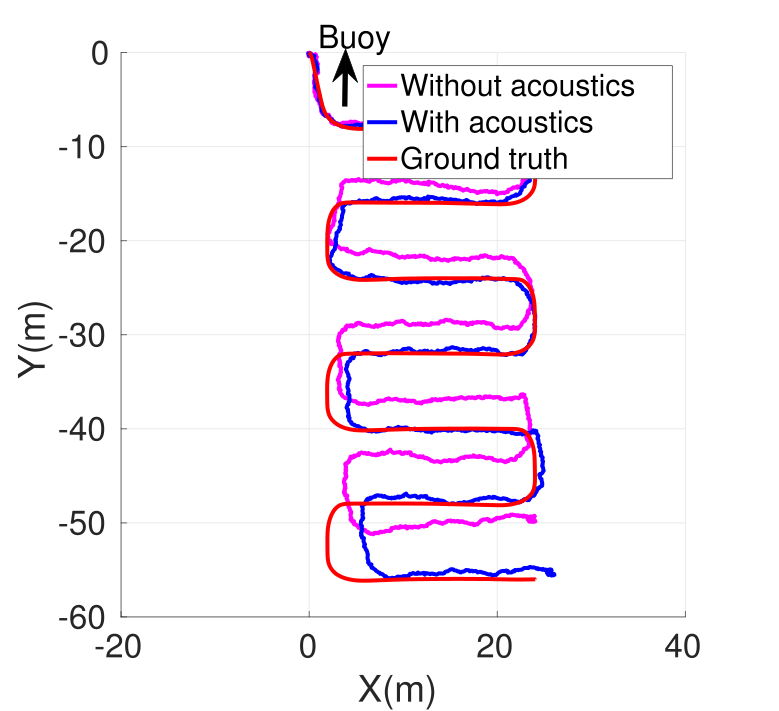}
}
\subfigure[\label{fig: lawn_2buoy_xy} Buoys placed at $(0, 50, 0)$ and $(75, -50, 0)$]{
\includegraphics[width=0.45\textwidth]{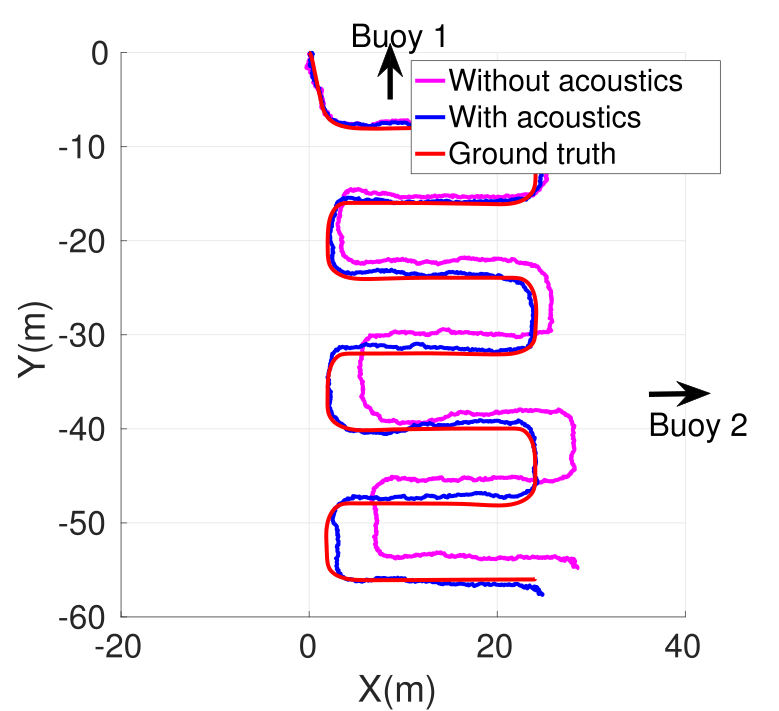}
}
\vspace{-4mm}
\caption[]{Simulation Results: Comparison of localization error between two  lawn mover surveys with one and two buoys. The trajectory of the vehicle without acoustic range sensors is shown in pink, while the range-assisted trajectory is shown in blue.}
\end{figure}

In this scenario the robot performs a survey of  sea floor. We start off from the surface and dive  to $1m$ before starting the survey path. The survey in itself consists of 7 track sequences of $20m$ length spaced at  a distance of $8m$.

In Figure~\ref{fig: lawn_seq} a single buoy transmitter is placed along the Y axis. Hence the estimate in the Y coordinate receives corrections while there still is significant drift along the X direction.
The same trajectory is repeated with an additional buoy, such that one performs corrections along the X axis while the other along Y axis. 
Figure\ref{fig: lawn_2buoy_xy} shows the improvement in the final trajectory using the two buoys, where errors are corrected in both X and Y. 

\subsubsection{Simulation results}

The X and Y errors are significantly more as compared to Z since the altimeter measures the depth of robot accurately. The results of the different scenarios are shown in Table \ref{tab:sim_results}.  While the use of a single range sensor shows improved performance as compared to raw odometry only, we get error corrections only along the radial direction. However in case of resource constrained survey missions using even one buoy can provide significant improvement to the estimation. 
Using 2 buoys on the other hand allows us to localize ourselves globally and also perform corrections both along the X and Y axes. Hence depending on the mission requirements one might choose to have 2 or more buoys for improved performance and redundancy.
\begin{table}
	\centering
	\caption{RMS Position Error for different simulation scenarios}
\begin{tabular}{lcc}
	\textbf{Scenario} & \textbf{Without Range sensor} & \textbf{With Range sensor} \\ \hline
	Dive & 10.04m & 2.85m\\
	Single transmitter Lawn Survey & 3.76m & 1.30m \\
	Two transmitter Lawn Survey &  3.56m & 1.04m\\
\end{tabular}
\label{tab:sim_results}
\end{table}
\vspace{-7mm}

\section{Conclusion}
\label{sec:conclusion}
This work has presented the development of a custom  acoustic ranging system.  Hardware transducers and the corresponding electronic filtering amplifiers  have been designed, assembled  and tested.  We have also shown how the use of range information from a single source produces a significant increase in pose estimation accuracy when fused together with low-cost vehicle sensors. Future work will focus on extending the range of the system, as well as integrating all data processing onto the embedded microcontroller. Further characterization of the delays introduced during the different processing steps will also enable to compensate the constant offset detected in the range estimates.   

\vspace{-2mm}
\section*{Acknowledgments}

The authors would like to thank Jim Trezzo for the work and research shared on the OpenRov forums, as well as Laura Giner for her help during the experimental data collection. The work presented has been partially supported by NASA award NNX16AL08G. 

\renewcommand{\bibfont}{\normalfont\small}
\printbibliography




\end{document}